# Modeling an Electrically Driven Graphene-Nanoribbon Laser for Optical Interconnects

Guangcun Shan[1], Chan-Hung Shek[1,*]
[1] Department of Physics and Materials Science, City University of Hong Kong, Hong Kong SAR
E-mail:guangcunshan@mail.sim.ac.cn;apchshek@cityu.edu.hk

*Abstract*— **Graphene has two very important optical properties of population inversion of electrons, and broadband optical gain. As a result, graphene has potential for use in lasers and amplifiers. In this work, we presented a quantum master model and analyzed the properties for the electrically pumped single-AGNR vertical-cavity surface-emitting lasers (VCSELs) to investigate the lasing action and laser properties for realistic experimental parameters. A semiclassical approximation for the output power and laser linewidth is also derived. The laser threshold power was several orders of magnitude lower than that currently achievable with semiconductor microlasers. Our results have demonstrated that a single-AGNR VCSEL can serve as a nanolaser with ultralow lasing threshold. Implementation of such a GNR-based VCSEL is especially promising for optical interconnection systems since VCSELs emit low optical power and single longitudinal mode over a wide wavelength spectral range through tailoring GNRs.**

## I. Introduction

In semiconductor lasers, vertical-cavity surface-emitting lasers (VCSELs) at around 1.3 µm have been expected to realize high-performance and low-cost light sources for fiber-optic communication systems [1],[2]. The large conduction band offset improves the temperature performance over that of conventional InP-based materials. The GaAs system provides high-performance AlGaAs–GaAs distributed Bragg reflector (DBR) mirrors and permits the use of the well-established oxide-confined GaAs-based VCSEL manufacturing infrastructure. VCSEL and smart pixel arrays are very important for constructing optical interconnection systems because they can emit, switch and process a number of broadband optical signals simultaneously [2]. Meanwhile, motivated by the desire to reduce the laser threshold, an ultimate microscopic limit of semiconductor lasers is a strongly coupled single quantum-emitter-cavity system [3]-[9]. In the past decade, considerable progress has been made in the fabrication of high-Q semiconductor microcavity with quasi-three-dimensional photon confinement to enable strong light–matter interaction [3]-[5]. In particular, a single-quantum-dot-cavity laser system with ultralow-threshold proposed as the solid-state analog to a thresholdless one-atom laser have been investigated for both fundamental interests and applications since 1999 [5]-[8].

At the same time, the facile fabrication of high-quality graphene-nanoribbons (GNRs) has stimulated extensive research since the successful fabrication and characterization of monolayer graphene [9]-[18]. Several previous studies have shown that the quasi-particle band gap of armchair-edged GNRs (AGNRs) is in the interesting energy range of 1-3 eV for 2-1 nm wide GNRs due to many-body effects [11]-[16]. Such findings stimulated the design of graphene-based applications in nanoscale optoelectronics. Moreover, demonstrations including high-speed photodetectors, optical modulators, plasmonic devices, and a microcavity-controlled graphene transistor have been reported so far [11]-[18]. It is worth pointing out that several experiments related to the quasiparticle band gap ($E_g$) in GNRs have proved the existence of finite band gap in GNRs [11],[13],[14].

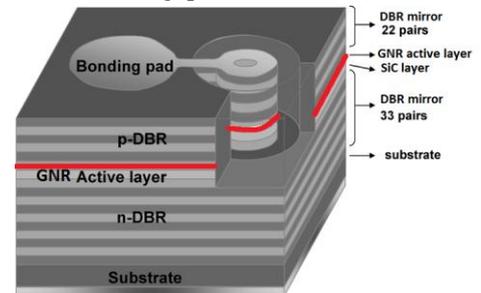

Fig. 1. Schematic of the configuration for the AGNR vertical cavity surface emitting laser (VCSEL) system. The active medium consists of a single-AGNR in a cavity matrix. The optical cavity is a DBR, which is brought close to the sample surface. Light emitted from the AGNR couples into a cavity mode of the DBR.

Nevertheless, little attention has been paid to the possibility that such a solid-state AGNR emitter as a novel active gain medium may replace the quantum dot gain emitter to achieve an ideal thresholdless laser. In light of these aforementioned issues, we consider the quantum master model of a coupled single-AGNR VCSEL device to examine whether such a device can be regarded as a thresholdless nanolaser by elucidating the fundamental quantum statistical properties of emitted photons. In Sec. II of this paper, we describe our proposed schematic structure and the corresponding quantum master model for an incoherently pumped single-AGNR VCSEL laser system. In Sec. III we calculate the steady state solutions of the mean photon number and the linewidth for realistic experimental parameters. Furthermore, we derived a semiclassical approximation for the linewidth in steady state. Besides, we also show how regular electrical pumping as a unique feature can be implemented in the single-AGNR VCSEL system.

## II. THEORY AND MODEL

As a specific illustration of the experimental setup shown in Fig.1, the GNR VCSEL system under study is a single AGNR coupled to a single mode of a DBR cavity. The active gain medium consisted of a single GNR on a Si substrate with SiC in between. Due to the E$g \propto 1/w$ relation, AGNR emitter with 1.3μm exciton emission (or with ~0.96eV band gap) can be realized by etching GNRs to suitable selective widths. Light emitted from the GNR couples into a cavity mode of the DBR. Notably, GNRs are excellent candidates for lasing gain emitter since they are found to be stable even without hydrogen passivation. In our model we assumed electrical pumping with unity internal quantum efficiency, ignoring effects such as leakage current. One way to approach this ideal limit is to incorporate the GNR into a double-heterojunction resonant-tunneling structure, similar to that used to inject excitons into quantum wells[19]. Carriers can tunnel from the doped Si reservoirs, through the intrinsic SiC barriers, into the single isolated AGNR. If an electron (hole) tunnels, a second electron (hole) requires a somewhat larger energy to tunnel due to the Coulomb repulsive energy. The electron (hole) energy level in the AGNR is effectively shifted by a single electron (hole) tunneling process. When the shifted energy level lies above the Fermi level of the n (p) side further electron (hole) tunneling is prohibited. This is known as the Coulomb blockade effect. Thus, only a single electron and single hole can tunnel into the electron and hole ground state of the GNR. This state, a single hole and a single electron in the AGNR, is the upper laser state |E>. The electron and hole radiatively recombine, leaving the GNR in the laser ground state |G>, i.e., no hole and no electron in the GNR. Moreover, the predominant lasing transition is from an upper (excited) state to a lower ground state, which is the lowest excited state having the main oscillator strength [1],[9],[17]. For this reason, we could take into account only this state for the exciton-photon interaction, and in the corresponding theoretical model, this exciton-photon coupling interaction is described by the Jaynes-Cummings Hamiltonian [1]-[4],[9]. On the other hand, we should notice that there are two paths by which to pump the AGNR to the upper laser state. Either a hole tunnels first, followed by an electron tunneling event, or an electron tunnels first, followed by a hole tunneling event. We denote the intermediate states |one hole>|zero electron> and |zero hole>|one electron> as states |J> and |I>, respectively. As a consequence, the system can be described as an incoherently pumped four-level laser as illustrated in Fig 2. Incoherent pumping of the upper laser level via intermediate levels |J> and |I> is described by rates $R_{GI}$, $R_{IE}$, $R_{GJ}$, and $R_{JE}$.

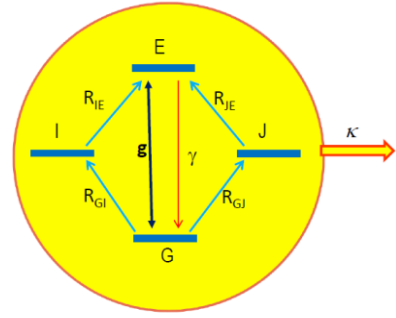

Fig. 2. Schematic level diagram of a single-AGNR VCSEL levels. The four levels are connected via rates $R_{GI}$, $R_{IE}$, $R_{GJ}$, and $R_{JE}$ (electron and hole tunneling). Spontaneous emission rate is described by $\gamma$, $g$ is the AGNR-field coupling constant, and $\kappa$ is the photon damping rate.

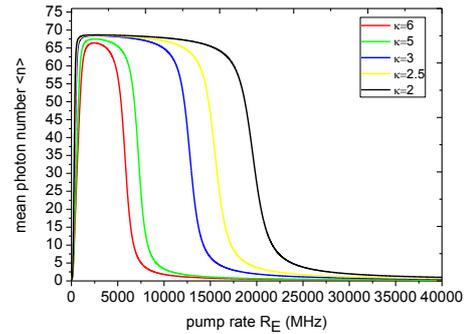

Fig. 3. Mean photon number <$n$> in the lasing mode versus excitation pump $R_E$ for different exciton-photon coupling strength $g$=200MHz. The photon damping rates $\kappa$ (from the bottom to the top) = 6(red), 5(green), 3(blue), 2.5(yellow), and 2 MHz(black), and $\gamma$=300MHz.

In the dipole approximation [20],[21], the Hamiltonian is further given by

$$H = \hbar\omega_c a^+ a + \hbar\omega_0 b^+ b + i\hbar g\left(a^+ b - ab^+\right), \quad (1)$$

where $\omega_c$ is the resonant frequency for cavity mode, $\omega_0$ is the corresponding lasing transition frequency of the exciton mode in AGNR that is assumed to be resonant with the cavity mode, and $g$ is the coupling strength of the transition between lasing levels, |E> and |G>, and the cavity mode of the DBR for the exciton-photon interaction. $a^+$ and $a$ are the boson creation and annihilation operators for photons of electromagnetic fields of cavity mode, respectively. $b^+$ =|E><G| and $b$=|G><E| are the atomic creation and annihilation operators for the exciton mode, respectively.

Applying the standard methods of the quantum theory of damping [3],[21],[22], the master equation in the general Lindblad form can be written as

$$\frac{\partial}{\partial t}\rho = L\rho = \frac{1}{i\hbar}[H,\rho] + L_{AGNR}\rho + L_{cavity}\rho + L_{pump}\rho \quad (2)$$

where $L_{AGNR}$ is the dissipation term corresponding to AGNR gain medium, $L_{cavity}$ is the cavity field dissipation term, and $L_{pump}$ is the incoherent pumping term. The dissipation terms

lead to non-unitary evolution given by

$$L_{AGNR} = -\frac{\gamma}{2}\left(|E\rangle\langle E|\rho + \rho|E\rangle\langle E| - 2|G\rangle\langle E|\rho|E\rangle\langle G|\right) \quad (3)$$

$$L_{cavity} = -\frac{\kappa}{2}\left(a^+a\rho + \rho a^+a - 2a\rho a^+\right) \quad (4)$$

where, the rate $\gamma$ describes the combined effect of spontaneous emission of photons in modes other than the lasing mode and nonradiative decay, and $\kappa$ is the photon damping rate of cavity in the laser mode. And incoherent pumping term is described by

$$\begin{aligned}L_{pump}\rho = &-\frac{R_{GI}}{2}\left(|G\rangle\langle G|\rho + \rho|G\rangle\langle G| - 2|I\rangle\langle G|\rho|G\rangle\langle I|\right)\\ &-\frac{R_{IE}}{2}\left(|I\rangle\langle I|\rho + \rho|I\rangle\langle I| - 2|E\rangle\langle I|\rho|I\rangle\langle E|\right)\\ &-\frac{R_{GJ}}{2}\left(|G\rangle\langle G|\rho + \rho|G\rangle\langle G| - 2|J\rangle\langle G|\rho|G\rangle\langle J|\right)\\ &-\frac{R_{JE}}{2}\left(|J\rangle\langle J|\rho + \rho|J\rangle\langle J| - 2|E\rangle\langle J|\rho|J\rangle\langle E|\right)\end{aligned} \quad (5)$$

where $R_{GI}$, $R_{IE}$, $R_{GJ}$, and $R_{JE}$ are the incoherent excitation pumping rate of the upper laser level for the AGNR via intermediate levels $|J\rangle$ and $|I\rangle$. Furthermore, the equations of motion for the projections of $\rho$ on the AGNR levels $\rho_{M,N} = \langle M|\rho|N\rangle$ ($M,N=E,G,I,J$) are readily obtained

$$\frac{\partial}{\partial t}\rho_{EE} = -ig(\rho_{EG}a^+ - a\rho_{GE}) + (L_{cavity} - R_{EG})\rho_{EE} \quad (6\text{-}1)$$
$$+ R_{IE}\rho_{II} + R_{JE}\rho_{JJ}$$

$$\frac{\partial}{\partial t}\rho_{GG} = -ig(\rho_{GE}a - a^+\rho_{EG}) \quad (6\text{-}2)$$
$$+ (L_{cavity} - R_{GI} - R_{GJ})\rho_{GG} + \gamma\rho_{EE}$$

$$\frac{\partial}{\partial t}\rho_{EG} = -ig(\rho_{EE}a - a\rho_{GG}) + (L_{cavity} - \Gamma)\rho_{EG} \quad (6\text{-}3)$$

$$\frac{\partial}{\partial t}\rho_{GE} = -ig(a^+\rho_{EE} - \rho_{GG}a^+) + (L_{cavity} - \Gamma)\rho_{GE} \quad (6\text{-}4)$$

$$\frac{\partial}{\partial t}\rho_{II} = (L_{cavity} - R_{IE})\rho_{II} + R_{GI}\rho_{GG} \quad (6\text{-}5)$$

$$\frac{\partial}{\partial t}\rho_{JJ} = (L_{cavity} - R_{JE})\rho_{JJ} + R_{GJ}\rho_{GG} \quad (6\text{-}6)$$

where $\Gamma = (\gamma + R_{GI} + R_{GJ})/2 + \gamma^*$, and $\gamma^*$ describes an additional pure dephasing rate, which is mainly due to phonon scattering. The quantum master equation, Eqn. (2) can be written as a first-order differential equations and solved by numerically using the Savage and Carmichael simulation method [5],[20],[21].

### III. RESULTS AND DISCUSSIONS

Now we first consider realistic parameters for our modeled electrically pumped GNR VCSEL system, and then examine the dynamics as well as the general steady state properties. In our modeled system in Fig. 1, we assume that the resonant-tunneling condition is satisfied simultaneously for electrons and holes. The electron and hole tunneling rates are determined by the design of the double-heterojunction structure, and depend on parameters such as tunnel barrier width and height, doping, and applied bias voltage. Based on the WKB approximation, it is concluded that the ratio of electron tunneling rate ($R_{GI}$ and $R_{JE}$) and hole tunneling rate ($R_{GJ}$ and $R_{IE}$) are having approximately equal value for designated equal tunnel barrier widths, because of the same effective mass of electrons and holes. If the pumping bias voltage is changed, all pump rates are swept simultaneously; thus, the relations $R_{GI}=R_{JE}=R_{GJ}=R_{IE}=R_E$ remain valid. In the following, we refer to $R_E$ (any one of the pumping rates $R_{ij}$) whenever we talk about the pump rate. The pump rates $R_E$ range from 500 MHz to over 10 GHz. We would like to point out that the assumption of equal constants for pump rates is not very stringent. However, the general properties of the system under consideration can be well described by the above assumptions as have demonstrated in the case of single-quantum-dot VCSEL, although it is straightforward to take into account a more general model, where the pump rates are controlled individually[3]. DBR cavities with Q values as high as $10^9$ can be made using novel semiconductor fabrication techniques [1],[2],[23]. In practice, Q values of $2.9\times10^8$, which corresponds to a photon damping rate of $\kappa$=5 MHz for 1.3μ nm lasing, can be readily achieved. The spontaneous decay time $\gamma$ for AGNR gain medium was estimated to be about 650 ps-15ns[9],[12]. Finally, in the system considered here, the emitter-cavity coupling constant $g$ can readily reach up to $g/2\pi$=33 MHz[3],[4],[9]. With these parameters, we solve Eq.(5) to investigate the properties of single AGNR VCSEL device.

Now, we investigate the steady-state mean photon number $\langle n \rangle = \langle n(t) \rangle_{ss} = \langle a^+(t)a(t) \rangle_{ss}$ versus pumping excitation rate $R_E$ for different values of coupling strength $g$. Fig. 3 shows the mean photon number in the lasing mode for different photon damping rates $\kappa$ as the pump rate is changed. The generic behavior for all $\kappa$ is characterized by an increase in the mean photon number up to a maximum value, followed by a monotonic decrease. This self-quenching is due to the destruction of the coherence between the lasing levels by the strong incoherent pump, which was also predicted for incoherently pumped single-quantum-dot microcavity lasers [4]-[6],[9]. The destruction of coherence can be interpreted in terms of quantum measurement. If a continuous measurement is performed to determine whether the AGNR gain medium is in state $E$ or $G$, the decay rate $\Gamma = (\gamma + 2R_E)/2$ of the coherences $\rho_{EG}$ and $\rho_{GE}$ in Eq. (6) can then be interpreted as the rate of individual measurements. At this point we should note that hereafter such an additional pure dephasing rate can be neglected without loss of rigor in our case of single-AGNR VCSEL device. It is indicated that increasing $g$ increases the intracavity mean photon number $\langle n \rangle$. It is also found that decreasing $\kappa$ also increases the intracavity mean photon number, which was also observed in one-atom laser[20],[21] and single-quantum-dot laser[3],[5],[8].

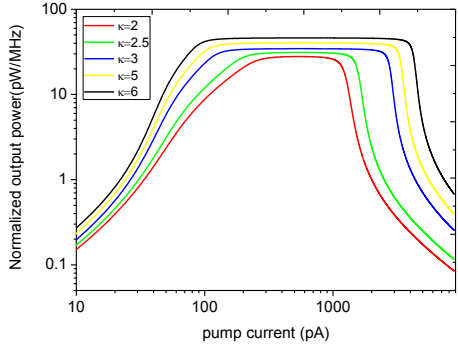

Fig. 4. Normalized output power $I$ (pW/MHz) versus excitation pumping current $R_E$ for a wider range of photon damping rates of $\kappa=6$ (black), 5(yellow), 3(blue), 2.5(green), and 2 MHz(red), respectively. The parameter $g=200$MHz. Note that the output power is normalized to $\kappa$ and the pumping rate is normalized to current.

Next, we evaluate the normalized output power $P_{out}$ to give a quantitative assessment of the performance of the single-AGNR VCSEL device for a wider range of photon damping rates $\kappa$. Fig. 5 shows the normalized output power $P_{out}/<\kappa>$ for $\kappa=6, 5, 3, 2.5,$ and 2 MHz, which corresponds to $Q=2.4\times10^8$, $2.9\times10^8$, $4.8\times10^8$, $5.8\times10^8$, and $7.3\times10^8$, respectively. We find that, above a photon number of unity in the lasing mode, a sharp increase in the photon number and decrease in the linewidth is observed as the pump strength $R_E$ is increased. The generic behavior for all $\kappa$ is characterized by an increase in the mean photon number up to a maximum value, followed by a monotonic decrease. This self-quenching was also predicted for incoherently pumped single-quantum-dot microcavity lasers due to the destruction of the coherence between the lasing levels by the strong incoherent pump [3],[5]. Moreover, the cavity Purcell effect leads to much more photons than those generated by transition via excitonic spontaneous decay of AGNR alone. The AGNR quantum lasing emitter is therefore essentially analogous to one-atom laser or single-quantum-dot microcavity laser [3],[8],[20],[21]. With increasing Q the onset of self-quenching is shifted to higher pump rates, and there is a larger regime with a linear input/output power characteristic. The threshold for $\kappa=5$ MHz corresponds to a current of 22.4 pA, which is more than five orders of magnitude lower than the current of 70μA record for a microcavity semiconductor laser [24]. The maximum output power of 35.6 pW in this case can be increased to several nW by increasing the Q of the advanced optical microcavity. From our simulation results, experimentally speaking, we infer that it may be possible to achieve a GNR surface-emitting nanolaser if $g>4\gamma$ and the cavity is a suitable good cavity ($\kappa<\gamma$).

Next, we mainly focus on the semiclassical approximation for the steady state properties of the single-AGNR VCSEL device. Since a laser is said to be at threshold semiclassically when the pumping rate is just sufficient for photon production to balance the losses, the semiclassical factorization of the density matrix is used for larger photon numbers for the steady state properties. The expectation values of atomic exciton operators and field operators are assumed to factorize, since the photon number here is large enough. Moreover, due to the short lifetime of the upper laser state in the system, the pump rates have to be large and the photon storage time to be long in order to achieve threshold. It is therefore a good approximation to neglect the field damping operator $L_{field}$ in the equation for the polarizations $\rho_{EG}$ and $\rho_{GE}$. At higher temperatures phonon scattering leads to a large dephasing rate $\gamma^*$, and the expectation value $<\rho_{EG}>$ can be adiabatically eliminated if the semiclassical factorization of the density matrix is used. As a result, the lasing transition for AGNR can be described as a two-level system with the ground (lower) state $|G>$ and the excited (upper) state $|E>$, which reduces Eq.(6) to an effective two-level system:

$$\langle\dot{n}\rangle = \frac{2g^2}{\Gamma}(\rho_{EE}-\rho_{GG})\langle n\rangle + \frac{2g^2}{\Gamma}\rho_{EE} - \kappa\langle n\rangle \quad (7\text{-}1)$$

$$\dot{\rho}_{EE} = -\frac{2g^2}{\Gamma}(\rho_{EE}-\rho_{GG})\langle n\rangle - \left(\frac{2g^2}{\Gamma}+\gamma\right)\rho_{EE} \quad (7\text{-}2)$$
$$+ 2R_E\rho_{GG}$$

$$\dot{\rho}_{GG} = \frac{2g^2}{\Gamma}(\rho_{EE}-\rho_{GG})\langle n\rangle + \left(\frac{2g^2}{\Gamma}+\gamma\right)\rho_{EE} \quad (7\text{-}3)$$
$$- 2R_E\rho_{GG}$$

where $\rho_{EE}$ and $\rho_{GG}$ are the probabilities of finding the AGNR in state $E$ and $G$, respectively, and $<n>$ is the mean photon number. We use the derived approximations for the linewidth to give a quantitative analysis of the steady state properties of the single-GNR VCSEL for a wider range of photon damping rates of $\kappa=6, 5, 3, 2.5,$ and 2 MHz, which corresponds to $Q=4.9\times10^8$, $6.5\times10^8$, $9.8\times10^8$, $1.9\times10^9$, and $3.9\times10^9$, respectively. Following the Yamamoto's approach described in Ref. [23], a semiclassical approximation for the linewidth is obtained as

$$\Delta\nu_{FWHM} = \frac{1}{2\pi}\left(\kappa - \frac{2g^2}{\Gamma}(\rho_{EE}-\rho_{GG})\right) \quad (8)$$

where $\rho_{EE}$ and $\rho_{GG}$ are the semiclassical steady state results for the effective two-level system. At small pump rates, the linewidth is broadened with respect to the empty cavity linewidth, due to the presence of the absorbing gain medium. As a result of the large coupling constant, this broadening is quite significant, even if the gain absorber is such a single AGNR quantum emitter. Above a photon number of unity in the lasing mode, a sharp decrease in the linewidth is observed. However, at very high pump rate, the AGNR is completely decoupled from the cavity mode due to self-quenching, and thus the linewidth broadens and approaches the empty cavity linewidth $\kappa/2\pi$ as the pump rate further increases.

We want to point out that in such a single-AGNR-cavity coupled system has promising properties for future applications such as single photon sources and low-cost light source for fiber-optic communication system [2],[25]. In the future, large-scale monolithic photonic integrated circuits (MPhICs) represent a significant technology innovation that simplifies optical communication system design, reduces space and power consumption, and improves reliability.

## IV. CONCLUSION

In summary, a theoretical quantum master model for the electrically pumped single-GNR VCSEL nanodevice has been presented in detail to investigate the laser properties and lasing action, such as output power, and mean photon number linewidth, for realistic experimental parameters. The laser threshold power is several orders of magnitude lower than that possible currently with semiconductor microlasers. When the single AGNR VCSEL device is electrically pumped, our theoretical results have demonstrated that a single-AGNR VCSEL can serve as a nanolaser with ultralow lasing threshold. Implementation of such a GNR-based VCSEL is especially promising for optical interconnection systems and graphene-based MPhICs since VCSELs emit low optical power and single longitudinal mode over a wide wavelength spectral range through tailoring GNRs.

## ACKNOWLEDGMENT

We would like to thank Prof. C. Cohen-Tannoudji for his fruitful discussions on quantum optics and nanolaser during his visit to City University of Hong Kong on 13 January 2011. This work was financially supported by a Strategic Research Grant (project number: 7008101 and 7002554) from City University of Hong Kong, Hong Kong SAR.